%% file: Dyncomm.tex
\documentclass[a4paper]{article}
\usepackage{arxiv}
\usepackage{enumerate}
\usepackage{amsmath}
\usepackage{graphicx}
\usepackage{multirow}
\usepackage{caption}
\usepackage{url}
\usepackage{hyperref}
\usepackage{float}
\usepackage[super]{nth}
\usepackage{url}
\usepackage[labelfont=bf]{caption}
\usepackage[labelsep=period]{caption}

\usepackage{algorithm} %ctan.org\pkg\algorithms
\usepackage{algpseudocode}

\usepackage{tikz}
\usetikzlibrary{backgrounds,fit,decorations.pathreplacing, shapes}  % TikZ libraries
\usepackage{tkz-graph}
\usepackage[utf8]{inputenc}
\usepackage[font=scriptsize,labelfont=scriptsize]{subfig}
\usepackage{subfig}

\usepackage[super]{nth}
\usepackage{adjustbox}
\usepackage{nameref}

\usepackage{listings}

\lstdefinestyle{mystyle}{
    basicstyle=\footnotesize,
    breakatwhitespace=false,         
    breaklines=true,                 
    captionpos=b,                    
    keepspaces=true,                 
    numbers=left,                    
    numbersep=5pt,                  
    showspaces=false,                
    showstringspaces=false,
    showtabs=false,                  
    tabsize=2
}
\algnewcommand{\LineComment}[1]{\State \(\triangleright\) #1}

\makeatletter
\def\@copyrightspace{\relax}
\makeatother

\usepackage[utf8]{inputenc}

\usepackage{tikz}
\usepackage{tkz-graph}
\usetikzlibrary{backgrounds,fit,decorations.pathreplacing, shapes}  % TikZ libraries
\usetikzlibrary{arrows, automata}
\usepackage{svg}
\usepackage{graphicx}
\usepackage{caption}
\usepackage{adjustbox}
\usepackage{multirow,multicol}
\usepackage{nameref}
\usepackage{nth}
\usepackage{url}
\usepackage{hyperref}
\usepackage{enumitem}
\usepackage{faktor}
\usepackage{scalerel,stackengine}

\makeatletter
\newcommand*\@dblLabelI {}
\newcommand*\@dblLabelII {}
\newcommand*\@dblequationAux {}

\def\@dblequationAux #1,#2,%
    {\def\@dblLabelI{\label{#1}}\def\@dblLabelII{\label{#2}}}

\newcommand*{\doubleequation}[3][]{%
    \par\vskip\abovedisplayskip\noindent
    \if\relax\detokenize{#1}\relax
       \let\@dblLabelI\@empty
       \let\@dblLabelII\@empty
    \else % we assume here that the optional argument
       \@dblequationAux #1,%
    \fi
    \makebox[0.5\linewidth-1.5em]{%
     \hspace{\stretch2}%
     \makebox[0pt]{$\displaystyle #2$}%
     \hspace{\stretch1}%
    }%
    \makebox[0.5\linewidth-1.5em]{%
     \hspace{\stretch1}%
     \makebox[0pt]{$\displaystyle #3$}%
     \hspace{\stretch2}%
    }%
    \makebox[3em][r]{(%
  \refstepcounter{equation}\theequation\@dblLabelI, 
  \refstepcounter{equation}\theequation\@dblLabelII)}%
  \par\vskip\belowdisplayskip
}
\makeatother

\makeatletter
\newcommand*\@trplLabelI {}
\newcommand*\@trplLabelII {}
\newcommand*\@trplLabelIII {}
\newcommand*\@trplequationAux {}

\def\@trplequationAux #1,#2,#3,%
    {\def\@trplLabelI{\label{#1}}\def\@trplLabelII{\label{#2}}\def\@trplLabelIII{\label{#3}}}

\newcommand*{\tripleequation}[4][]{%
    \par\vskip\abovedisplayskip\noindent
    \if\relax\detokenize{#1}\relax
       \let\@trplLabelI\@empty
       \let\@trplLabelII\@empty
       \let\@trplLabelIII\@empty
    \else % we assume here that the optional argument
       \@trplequationAux #1,%
    \fi
    \makebox[0.333\linewidth-1.5em]{%
     \hspace{\stretch2}%
     \makebox[0pt]{$\displaystyle #2$}%
     \hspace{\stretch1}%
    }%
    \makebox[0.333\linewidth-1.5em]{%
     \hspace{\stretch1}%
     \makebox[0pt]{$\displaystyle #3$}%
     \hspace{\stretch1}%
    }%
    \makebox[0.333\linewidth-1.5em]{%
     \hspace{\stretch1}%
     \makebox[0pt]{$\displaystyle #4$}%
     \hspace{\stretch2}%
    }%
    \makebox[3em][r]{(%
  \refstepcounter{equation}\theequation\@trplLabelI, 
  \refstepcounter{equation}\theequation\@trplLabelII,
  \refstepcounter{equation}\theequation\@trplLabelIII)}%
  \par\vskip\belowdisplayskip
}
\makeatother

\usepackage{amsmath}
\usepackage{mathptmx}
\usepackage{amssymb}
\usepackage{amstext}
\usepackage{amsthm}

\DeclareMathAlphabet{\mathcal}{OMS}{cmsy}{m}{n}

\tikzstyle{vertex}=[circle,draw=black, fill=white,sloped,minimum size=17pt,inner sep=5pt]
\newcommand{\vertex}{\node[vertex]}

\usepackage{tikz}
\usetikzlibrary{decorations.pathreplacing,calc}

\newcommand\MyBox[1]{
  \fbox{\lower0.75cm
    \vbox {\vfil
      \hbox {\hfil{#1}\hfil}
      \vfil}%
  }%
}

\newcommand{\figref}[1]%
{Figure \ref{#1}%
}

\newcommand{\tableref}[1]%
{Table \ref{#1}%
}

\newcommand{\algorithmref}[1]%
{Algorithm \ref{#1}%
}

\newcommand{\sectionref}[1]%
{Section \ref{#1}%
}

\newcommand{\lineref}[1]%
{Line \ref{#1}%
}

\begin{document}
%\long\def\/*#1*/{}

\title{DynComm R Package - Dynamic Community Detection for Evolving Networks}

\author{Rui Portocarrero Sarmento \\
PRODEI - Faculty of Engineering, UP\\
Rua Dr. Roberto Frias, s/n\\
4200-465 Porto, Portugal\\
mail@ruisarmento.com \And
Luís Lemos\\
FEUP - Faculty of Engineering, UP\\
Rua Dr. Roberto Frias, s/n\\
4200-465 Porto, Portugal\\
\And
Mário Cordeiro\\
PRODEI - Faculty of Engineering, UP\\
Rua Dr. Roberto Frias, s/n\\
4200-465 Porto, Portugal\\
\And
Giulio Rossetti \\
ISTI-CNR \\
Via Giuseppe Moruzzi, 1, \\
56127 Pisa PI, Italy \\
giulio.rossetti@isti.cnr.it
\And
Douglas Cardoso \\
CEFET-RJ \\
Petrópolis, Brazil \\
douglas.cardoso@cefet-rj.br
}
\maketitle

\begin{abstract}

Nowadays, the analysis of dynamics in networks represents a great deal in the Social Network Analysis research area. To support students, teachers, developers, and researchers in this work we introduce a novel R package, namely DynComm. It is designed to be a multi-language package, that can be used for community detection and analysis on dynamic networks. The package introduces interfaces to facilitate further developments and the addition of new and future developed algorithms to deal with community detection in evolving networks. This new package has the goal of abstracting the programmatic interface of the algorithms, whether they are written in R or other languages, and expose them as functions in R.


\end{abstract}

%\section{Introduction}\label{sec:intro}
%\tableofcontents

\section{Introduction}

Community Detection in Social Network Analysis (SNA) is a critical research area in an enormous amount of unrelated areas. From Psychology to Physics, community detection in SNA is used to find an agglomeration of objects of study in a graph. We have witnessed the development of an abundance of algorithms specifically designed to identify communities in graphs of every size, from small graphs with some dozens of vertices and edges, to large or very large graphs with millions of vertices and billions of edges.

Very recently, with the growing popularity of SNA, researchers have been migrating concepts and some algorithms to stream approaches. This is even more important with the appearance of sources of data that are streamable, such as social networks like Facebook and Twitter, where information arrives as a flow of discrete events that, usually, have a limited existence through time.

Available since some years ago, a variety of packages in languages as Python or R have been developed to cope with the need for analysis of communities and social networks. Nonetheless, few or no packages that deal with community detection in evolving networks are available right now, in R-CRAN. Thus, this is the right time to develop a way to provide researchers with a package that fulfills the need to explore network streams, particularly community detection of evolving networks. This poses a challenge since the adaptation of algorithms is not an easy task, and sometimes even impossible due to restrictions in the architecture of the static algorithm.

We tried to develop a framework that we believe will help future developments in this area to be included in the package, with the least amount of effort for the new algorithms' authors. 

The paper is organized as follows. In Section \ref{FromStatictoEvolvingNetworks}, we introduce some concepts related to evolving networks. Then, in section \ref{Dyncomm}, we explore the developed package and list its features, in the present version of development. In section \ref{INT}, we explain the developed R interface, in more detail. Then, in section \ref{API}, we explain the steps needed to add a new algorithm developed in other languages. Finally, in section \ref{conc}, we conclude this document by suggesting further developments of the DynComm R package \footnote{Available Code at \url{https://github.com/softskillsgroup/DynComm-R-package}.}.

\section{From Static to Evolving Networks}\label{FromStatictoEvolvingNetworks}
    
\begin{figure}
\hfill
\subfloat[Evolving Network]{
    \includegraphics[width = .495\columnwidth]{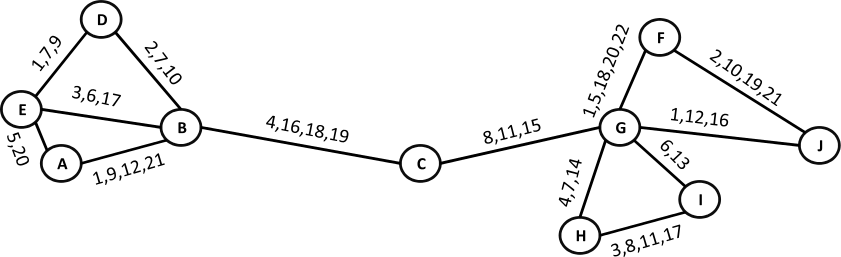}
\label{fig:TemporalNet:fig1}}
\hfill
\subfloat[Timeline Plot]{
    \includegraphics[width = .35\columnwidth]{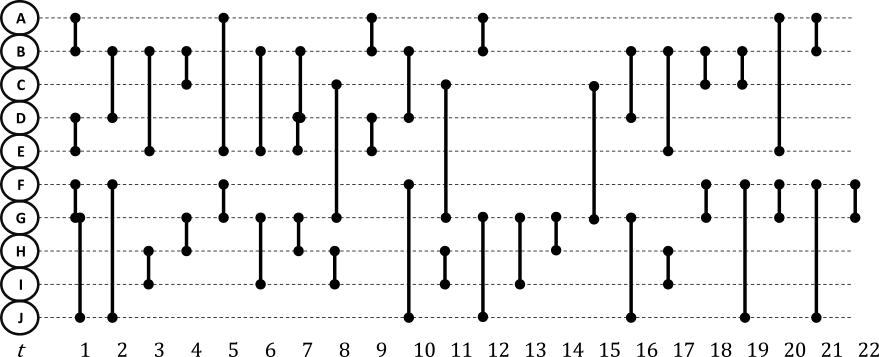}
\label{fig:TemporalNet:fig2}}
\hfill
\caption{Example of contact Evolving Network: \figref{fig:TemporalNet:fig1} shows a labelled aggregate network where the labels denote the times of contact, and \figref{fig:TemporalNet:fig2} shows a time-line plot, where each of the lines corresponds to one vertex and time runs from left to right}
\label{fig:TemporalNet}
\end{figure}

\begin{figure}
\hfill
\subfloat[Evolving Network]{
    \includegraphics[width = .495\columnwidth]{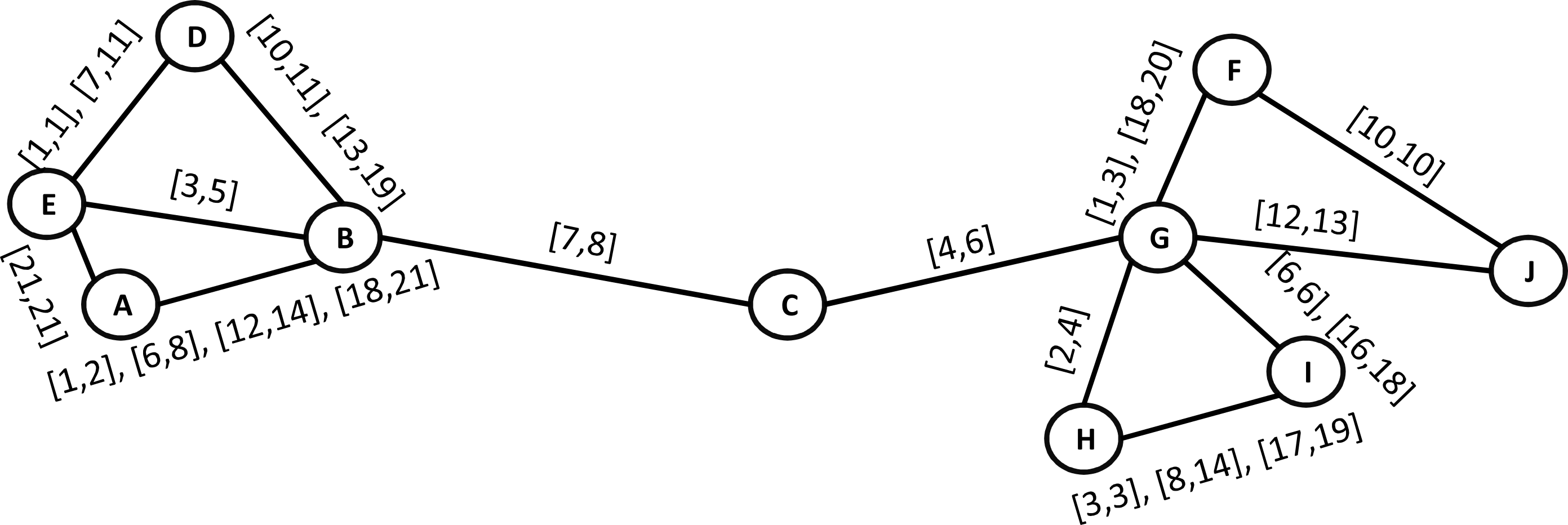}
\label{fig:TemporalNet2:fig1}}
\hfill
\subfloat[Timeline Plot]{
    \includegraphics[width = .35\columnwidth]{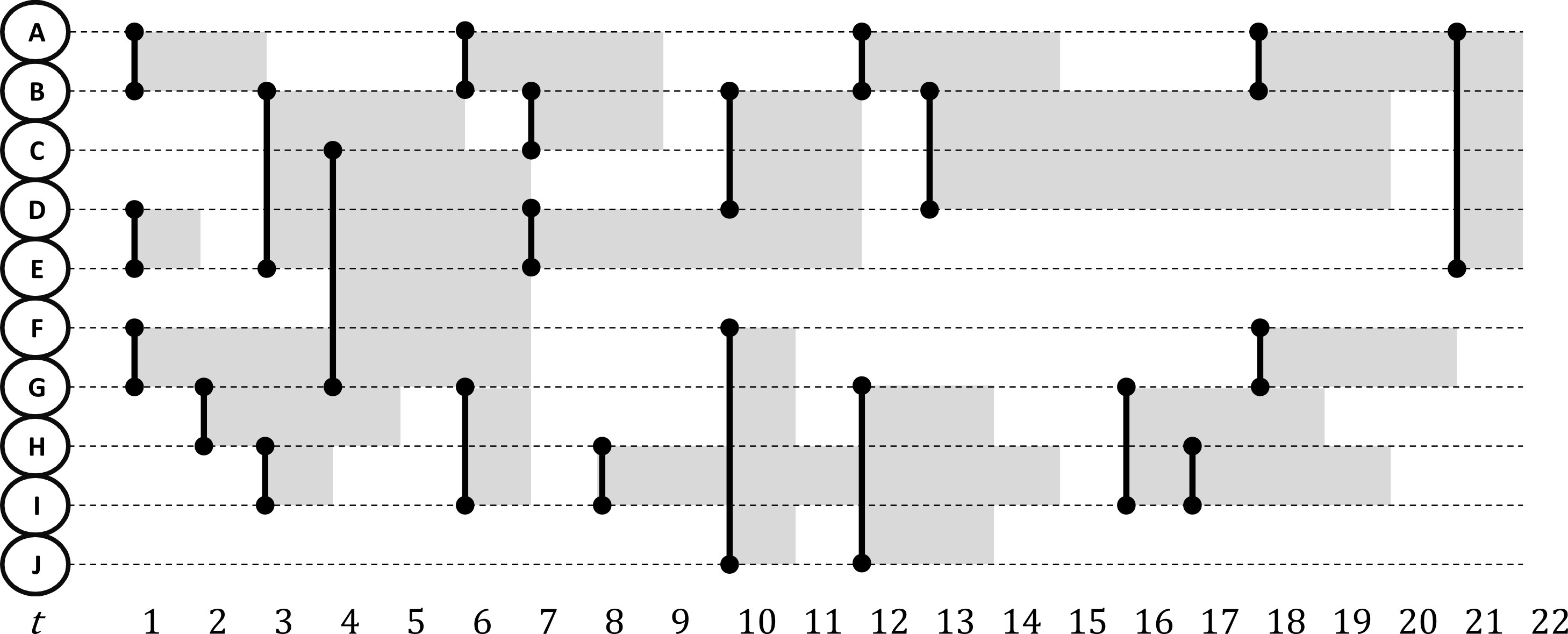}
\label{fig:TemporalNet2:fig2}}
\hfill
\caption{Example of interval Evolving Network: \figref{fig:TemporalNet2:fig1} shows the labelled aggregate network where the labels denote the time interval of the relation, and \figref{fig:TemporalNet2:fig2} shows a time-line plot, where each of the lines corresponds to one vertex and grey zones the time duration between two edges}
\label{fig:TemporalNet2}
\end{figure}
From \citep{cordeiro2018evolving}, \figref{fig:TemporalNet} shows the example of a contact Evolving Network with instantaneous interactions between vertices. When the interaction between network peers has a time duration we are in the presence of interval Evolving Networks as shown in \figref{fig:TemporalNet2:fig2}. Assuming that the time $T$ during which a network is observed is finite we can consider the start point $t_{start} = 0$ and the end time as $t_{end} = T$. A dynamic network graph $G_{0,T}^{D}(V, E_{0,T})$ on a time interval $[0,T[$ consists of a set of vertices $V$ and a set of temporal edges $E_{0,T}$. The evolving network is a set of graphs across the time axis within discrete time points $t_1, t_2, ..., t_{n-1}, t_n$. At time point $t_n$ is observed a graph instance $G(V_n, E_n)$ also denoted as $G_n$ where $E_n$ is the set of temporal edges $(u,v)_{t_n} \in E_{0,T}$ at time point $t_n$ with edges between vertices $u$ and $v$ on time interval $t_n = [t_{n_{begin}},t_{n_{end}}]$ such that $t_{n_{begin}} \le T$ and $t_{n_{end}} \ge t_{n_{begin}} \ge 0$. Examples of network changes that may occur between two time points $t_{n-1}$ and $t_{n}$ are the addition of new edges, i.e.: $E_n \supset E_{n-1}$, and the appearance of additional vertices, i.e.: $V_n \supset V_{n-1}$. 

\subsection{Models of Temporal Representation}
\label{FromStatictoEvolvingNetworks:ModelsofTemporalRepresentation}

%add Time-Varying Graphs model by Santoro, N., Quattrociocchi, W., Flocchini, P., Casteigts, A., & Amblard, F. (2011). Time-Varying Graphs and Social Network Analysis: Temporal Indicators and Metrics. Retrieved from http://arxiv.org/abs/1102.0629

%add time-ordered graph G model by Kim, H., & Anderson, R. (2012). Temporal node centrality in complex networks. Physical Review E, 85(2), 26107. http://doi.org/10.1103/PhysRevE.85.026107

%add time-graphlet or t-graphlet model by Thompson, W. H. H., Brantefors, P., & Fransson, P. (2016). From static to temporal network theory - applications to functional brain connectivity. bioRxiv, 96461. http://doi.org/10.1101/096461

\begin{figure}
\hfill
\subfloat[Aggregated]{
    \includegraphics[width = .120\columnwidth]{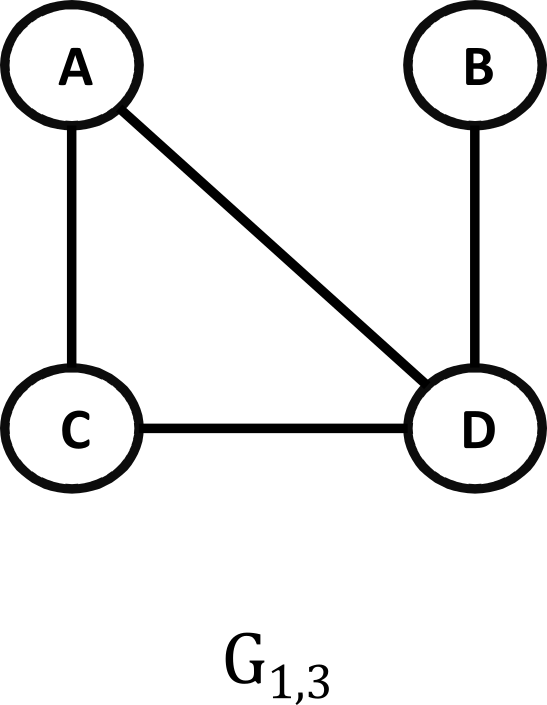}
\label{fig:TimeOrderedGraph:fig1}}
\hfill
\subfloat[Time-varying]{
    \includegraphics[width = .475\columnwidth]{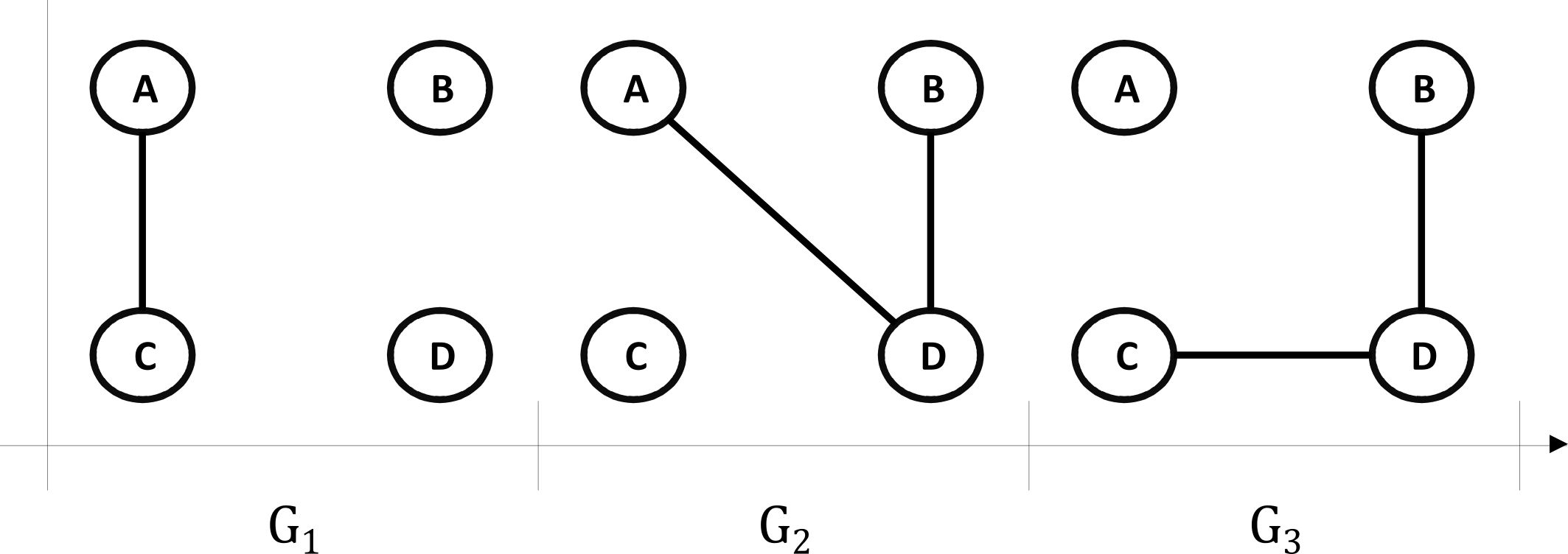}
\label{fig:TimeOrderedGraph:fig2}}
\hfill
\subfloat[Time-ordered]{
    \includegraphics[width = .295\columnwidth]{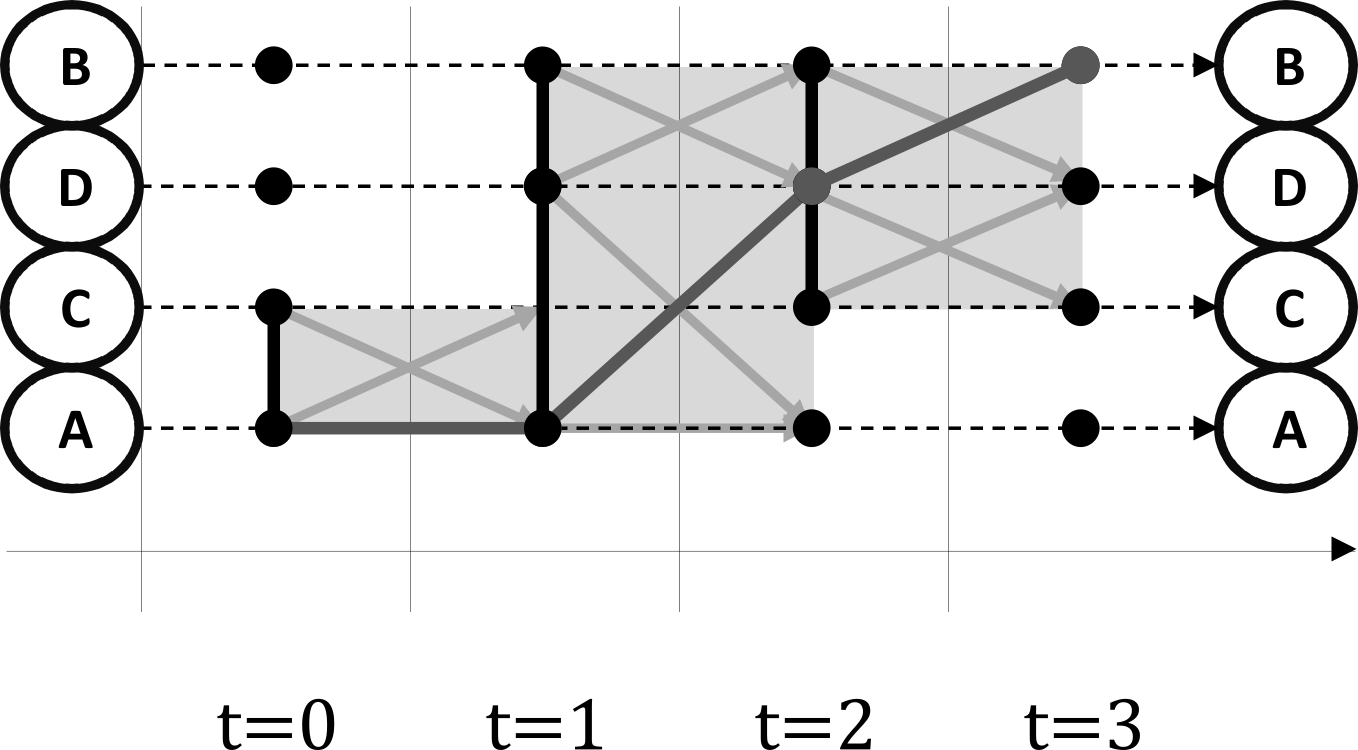}
\label{fig:TimeOrderedGraph:fig3}}
\hfill
\caption{Comparison of aggregated representation (\figref{fig:TimeOrderedGraph:fig1}) and time series representation (\figref{fig:TimeOrderedGraph:fig2}). The corresponding time ordered graph $\mathcal{G}$ is presented for the interval $[0,3]$ (\figref{fig:TimeOrderedGraph:fig3}).}
\label{fig:TimeOrderedGraph}
\end{figure}

\figref{fig:TimeOrderedGraph} presents the concept of a time-ordered graph for an example network for the time interval $[0,3]$. \figref{fig:TimeOrderedGraph:fig1} shows all the time intervals aggregated into a single graph $G_{1,3}$. The discretization of the network by converting the temporal information into a sequence of $n$ snapshots is presented in \figref{fig:TimeOrderedGraph:fig2}. In this example the evolving network is represented as a series of static networks $G_1,G_2, ... ,G_n$. The time-ordered graph  $\mathcal{G} = (\mathcal{V},\mathcal{E})$ of \figref{fig:TimeOrderedGraph:fig3} assumes that at each time step, a message can be delivered along a single edge. In the example of \figref{fig:TimeOrderedGraph:fig3} we show the temporal shortest path from vertex $u = A$ to vertex $v = B$. The temporal shortest path from $A$ to $B$ in the interval $[0,3]$ is $A_0 \rightarrow A_1 \rightarrow D_2 \rightarrow B_3$. The time-ordered graph of \citep{Kim2012} is the model used in the rest
of this document and package.

\subsection{Landmark vs Sliding Windows}
\label{FromStatictoEvolvingNetworks:LandmarkvsSlidingWindows}

\begin{figure}[h]
\hfill
\subfloat[Landmark]{
    \includegraphics[width = .275\columnwidth]{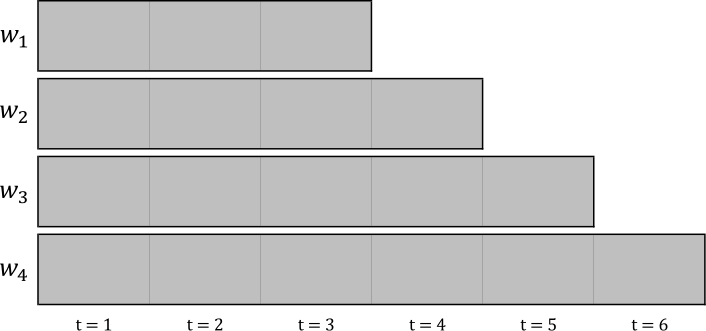}
\label{fig:TypesDataWindows:fig1}}
\hfill
\subfloat[Non-overlapping Sliding]{
    \includegraphics[width = .275\columnwidth]{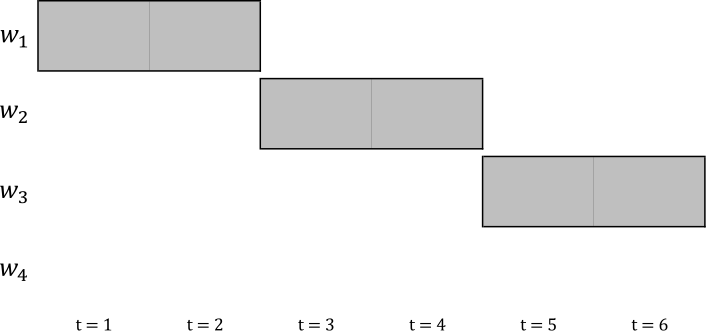}
\label{fig:TypesDataWindows:fig2}}
\hfill
\subfloat[Overlapping Sliding]{
    \includegraphics[width = .275\columnwidth]{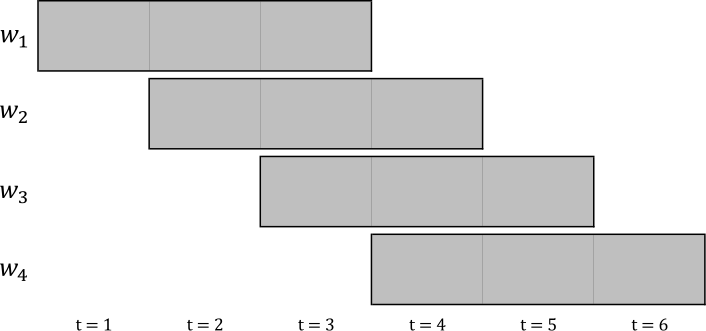}
\label{fig:TypesDataWindows:fig3}}
\hfill
\caption{Types of Data Windows: Landmark Window (\figref{fig:TypesDataWindows:fig1}) Non-overlapping Sliding Window (\figref{fig:TypesDataWindows:fig2}) and Overlapping Sliding Window (\figref{fig:TypesDataWindows:fig3})}
\label{fig:TypesDataWindows}
\end{figure}

When the temporal dimension is added to the analysis of networks, methodologies relating to the strategy to deal with knowledge that is being analyzed vary. \figref{fig:TypesDataWindows} shows 3 kinds of graph knowledge windowing ways. \textbf{\textit{Landmark windows}} \citep{Gehrke2001} comprehend all the info from a particular purpose in time, up to the present moment. within the Landmark window, the model is initialized in an determined point in time, i.e., the landmark that marks the start of the window. In ordered snapshots, the info window grows to think about all the data seen up to now, since the landmark start. \textbf{\textit{Sliding windows}}, from other point of view, are appropriate when we are not inquisitive about computing statistics over all the past, solely over the recent past \citep{Gama:2010:KDD:1855075}. \citep{Datar2002} incorporates a forgetting mechanism, does not take into account all the data falling outside the window, by keeping solely the newest data within the window. These windows can be defined regarding length in two distinct ways, the time-based length and the sequence-based \citep{Babcock:2002:SMW:545381.545465, Babcock:2002:MID:543613.543615}. Sequence-based models, wherever the dimensions of the window are, is set relating to the amount of observations. In Timestamp-based models, the other type of window generation, the dimensions of the window is outlined regarding time sample length. A timestamp window of size $t$ consists of all event elements whose timestamp is within a time interval $t$ since the beginning of the data processing, or since the beginning of the current period of processed data.

The implementation currently available in our package is explained by \ref{fig:TypesDataWindows:fig2}.

\subsection{Dynamic Community Detection}\label{CommunityDetection}

As a consequence of both global and local heterogeneity of edge distribution in a graph, specific regions of a graph evidence high concentration of edges within particular regions, called \textit{communities}, whereas inter regions have low concentrations of edges. In the context of networks, these occurrences of groups of vertices in a network that are more densely connected internally than with the rest of the network is called \textit{community structure}. Also known as \textit{modules} or \textit{clusters}, communities can, therefore, be straightforwardly defined as groups of similar vertices. A complete definition using the concept of density can be the following: communities can be understood as densely connected groups of vertices in the network, with sparser connections between them.

\subsubsection{Finding Communities in Static Networks}\label{CommunityDetection:FindingCommunitiesinStaticNetworks}

\begin{figure}[t!]
\centering
\subfloat[original network]{\input{plots/louvain/01}\label{fig:AddLouvain:addlouvain01}}\quad
\subfloat[initial communities]{\input{plots/louvain/02}\label{fig:AddLouvain:addlouvain02}}\quad
\subfloat[step 1 of \nth{1} iteration]{\input{plots/louvain/03}\label{fig:AddLouvain:addlouvain03}}\quad\\
\subfloat[step 2 of \nth{1} iteration]{\input{plots/louvain/04}\label{fig:AddLouvain:addlouvain04}}\quad
\subfloat[step 1 of \nth{2} iteration]{\input{plots/louvain/05}\label{fig:AddLouvain:addlouvain05}}\quad
\subfloat[step 2 of \nth{2} iteration]{\input{plots/louvain/06}\label{fig:AddLouvain:addlouvain06}}\quad
\caption{Example of an agglomerative community detection algorithm. In this case the original Louvain \citep{Blondel2008} with all algorithm steps.}
\label{fig:AddLouvain}
\end{figure}
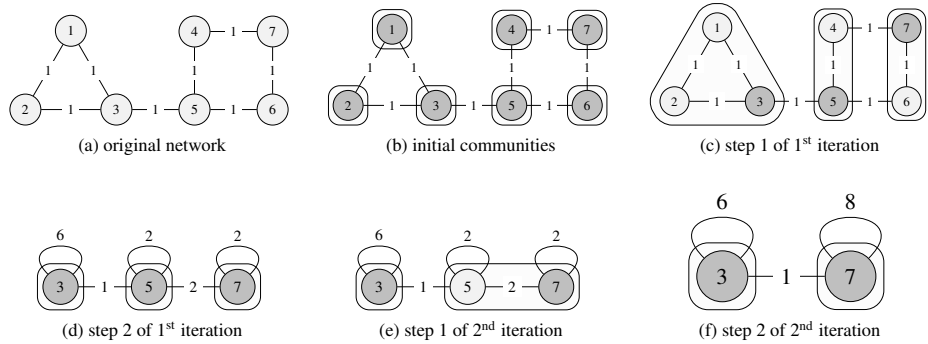

A greedy algorithm based on modularity optimization has been introduced by \citep{Blondel2008} where initially all vertices of the graph are put in different communities (\figref{fig:AddLouvain:addlouvain02}). The first step consists of a sequential sweep over all vertices, for each of the neighbors picks the community that yields the largest increase of modularity (\figref{fig:AddLouvain:addlouvain03}). At the end of the sweep, one obtains the first level partition. In the second step, communities are replaced by super vertices, and weight of the edge between the super vertices is the sum of the weights of the edges between the represented communities at the lower level (\figref{fig:AddLouvain:addlouvain04}). The two steps of the algorithm are then repeated, yielding new hierarchical levels and supergraphs (\figref{fig:AddLouvain:addlouvain06}).

\subsubsection{Finding Communities in Dynamic Networks}\label{CommunityDetection:FindingCommunitiesinDynamicNetworks}

When discussing methods for finding communities in dynamic networks the division of methods for slowly evolving networks and streaming networks is consensual \citep{Aggarwal2014}. In the following section, algorithms for both scenarios will be presented and analyzed

\paragraph{Slowly Evolving Networks}
\label{CommunityDetection:FindingCommunitiesinDynamicNetworks:SlowlyEvolvingNetworks}

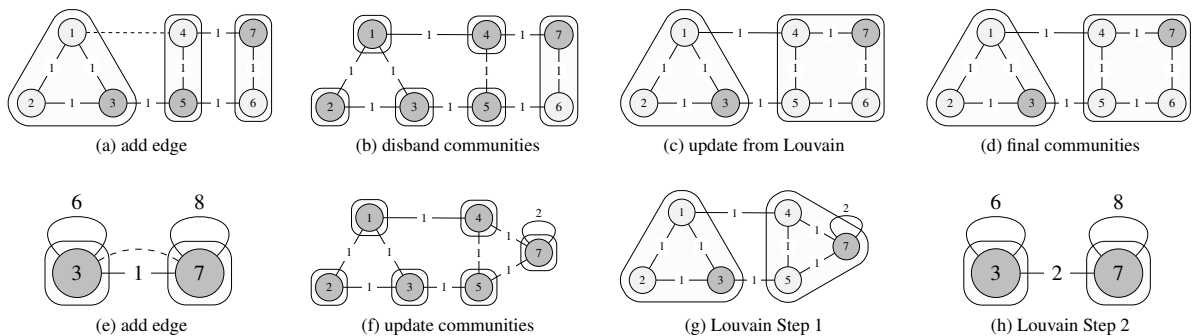
\begin{figure}
  \centering
    \subfloat[][add edge]{\input{plots/dlouvain/toynet05}\label{fig:AddCrossCommunityEdge:addstep05}}
    \quad
    \subfloat[][disband communities]{\input{plots/dlouvain/toynet06}\label{fig:AddCrossCommunityEdge:addstep06}}
    \quad
    \subfloat[][update from Louvain]{\input{plots/dlouvain/toynet09}\label{fig:AddCrossCommunityEdge:addstep09}}
    \quad
    \subfloat[][final communities]{\input{plots/dlouvain/toynet09}\label{fig:AddCrossCommunityEdge:addstep09h}}
    \quad
    \subfloat[][add edge]{\input{plots/dlouvain/toynet05h}\label{fig:AddCrossCommunityEdge:addstep05h}}
    \quad
    \subfloat[][update communities]{\input{plots/dlouvain/toynet07}\label{fig:AddCrossCommunityEdge:addstep07}}
    \quad
    \subfloat[][Louvain Step 1]{\input{plots/dlouvain/toynet08}\label{fig:AddCrossCommunityEdge:addstep08}}
    \quad
    \subfloat[][Louvain Step 2]{\input{plots/dlouvain/toynet10}\label{fig:AddCrossCommunityEdge:addstep10}}
    
\caption{Example of \citep{cordeiro2016dynamic} Dynamic Louvain for the addition of a Cross Community Edge (1 --- 4). Top figures show the lower-level network. At the bottom, are shown, the corresponding upper-level network with aggregated communities.}
\label{fig:AddCrossCommunityEdge}
\end{figure}

\citep{cordeiro2016dynamic} presented a modularity-based dynamic community detection algorithm. It is a modification of the original Louvain method where dynamically added and removed vertices and edges only affect their related communities. In each iteration, all the communities that were not affected by modifications to the network maintain unchanged. By reusing community structure obtained by previous iterations, the local modularity optimization step operates in smaller networks. Thus, only affected communities are disbanded to their origin. Compared with the original algorithm (\figref{fig:AddCrossCommunityEdge}), the stability of communities is also an improvement. When compared with the original algorithm, with that algorithm run several times, the results in changes on communities or vertex drift from one community to another are easier to follow, when presented by the dynamic and incremental algorithm. This is due to the fact that only parts of the network change during iterations, the non-determinism of the algorithm will have reduced effect on the community assignment, providing better community stability than its counterparts.

\begin{algorithm}[th]
  \footnotesize
  \caption{Dynamic Community Detection Algorithm}\label{DynamicCommunityDetectionAlgorithm}
  \begin{algorithmic}[1]
    \State $V\gets \{u_1,u_2, ..,u_v\}$ ,  $E\gets \{(i_1,j_1),(i_2,j_2), ..,(i_e,j_e)\}$          

    \State $A\gets array\{(i_1, j_1), .., (i_m, j_m)\}$             
    \State $R\gets array\{(i_1, j_1), .., (i_n, j_n)\}$             
    
    \Procedure{Main}{$G\gets (V,E),A,R$}                                                    \label{lst:line:Main}                           
    \State$\mathcal{C}_{ll} \gets \{C_1, C_2, .., C_n\}$, $\mathcal{C}_{ul} \gets \{\}$, $\mathcal{C}_{aux} \gets \mathcal{C}_{ll}$                
    
    \State $\Call{InitPartition}{\mathcal{C}_{aux} }$                                       \label{lst:line:InitPartition}        
    \State $mod \gets \Call{Modularity}{\mathcal{C}_{aux} }$, $old\_mod \gets 0$            \label{lst:line:Modularity1} 
    
    \State $m \gets 1$, $n \gets 1$
    
    %Main Loop%

    \While{$(mod \ge old\_mod \lor m \le \vert A \vert \lor v \le \vert R \vert )$}
      
      \State $\mathcal{C}_{aux}\gets \Call{OneLevel}{\mathcal{C}_{aux}}$ \label{lst:line:OneLevel}
    
      \State $\langle n, c \rangle \gets \Call{CommunityChangedVertices}{\mathcal{C}_{ll}, \mathcal{C}_{aux}}$ \label{lst:line:CommunityChangedVertices}
      
      \State $\mathcal{C}_{ll} \gets \Call{UpdateCommunities}{\mathcal{C}_{ll}, n, c}$        \label{lst:line:UpdateCommunities}      
      \State $old\_mod \gets mod$, $mod \gets \Call{Modularity}{\mathcal{C}_{ll} }$             \label{lst:line:Modularity2}                          
      
      \State$\mathcal{C}_{ul} \gets \Call{PartitionToGraph}{\mathcal{C}_{ll}}$             \label{lst:line:PartitionToGraph}             
      
      %Add New Edges%
      \If {$m \le \vert A \vert$}                                                                   
        \State $\langle src, dest\rangle \gets A[m]$                                                
        \State $a_{vertices} \gets \Call{AffectedByAddition}{src, dest, \mathcal{C}_{ll}}$ \label{lst:line:AffectedByAddition}
        \State $\mathcal{C}_{ll} \gets \Call{AddEdge}{src, dest, \mathcal{C}_{ll}}$     \label{lst:line:AddEdge}           
        \State $\mathcal{C}_{ll} \gets \Call{DisbandCommunities}{\mathcal{C}_{ll}, a_{vertices}}$     \label{lst:line:DisbandCommunitiesAdd}             
        \State $\mathcal{C}_{ul} \gets \Call{SyncCommunities}{\mathcal{C}_{ll}, \mathcal{C}_{ul}, a_{vertices}}$ \label{lst:line:SyncCommunitiesAdd}   
      \EndIf
      
      %Remove Edges%
      \If {$n \le \vert R \vert$}                                                                   
        \State $\langle src, dest\rangle \gets R[n]$                                                
        \State $a_{vertices} \gets \Call{AffectedByRemoval}{src, dest, \mathcal{C}_{ll}}$ \label{lst:line:AffectedByRemoval}
        \State $\mathcal{C}_{ll} \gets \Call{RemoveEdge}{src, dest, \mathcal{C}_{ll}}$   \label{lst:line:RemoveEdge}           
        \State $\mathcal{C}_{ll} \gets \Call{DisbandCommunities}{\mathcal{C}_{ll}, a_{vertices}}$      \label{lst:line:DisbandCommunitiesRem}          
        \State $\mathcal{C}_{ul} \gets \Call{SyncCommunities}{\mathcal{C}_{ll}, \mathcal{C}_{ul}, a_{vertices}}$ \label{lst:line:SyncCommunitiesRem}  
      \EndIf
      \State $\mathcal{C}_{aux} \gets \mathcal{C}_{ul}$, $m \gets m + 1$, $n \gets n + 1$ 
      
    \EndWhile

  \EndProcedure
  
\end{algorithmic}
\end{algorithm}

\paragraph{Streaming Networks}
\label{CommunityDetection:FindingCommunitiesinDynamicNetworks:StreamingNetworks}

Streaming graph algorithms are essential to perform community detection with high frequency data and large or very large networks. 
In streaming scenarios, the ability to perform deletion of edges in community detection algorithms is important. In short, as discussed in \sectionref{FromStatictoEvolvingNetworks:LandmarkvsSlidingWindows}, this will dictate if the method of analysis is to be performed over a sliding window of edges, and therefore edges are deleted from the tail end of the sliding window, or over a landmark window, in case there is no possibility to delete or forget old edges. Several methods were proposed for dynamic community discovery in graph streams. \citep{Wang2013} motivated by the variability of the underlying social behavior of individuals over different graph regions modeled the problem according to the so-termed \textit{local heterogeneity}, where a Local Weighted-Edge-based Pattern (LWEP) summary is efficiently maintained and used afterward to cluster the graph stream and perform dynamic community detection in weighted graph streams. Taking an almost linear time, \citep{Raghavan2007} investigated and a simple label propagation algorithm that uses the network structure alone as its guide and requires neither optimization of a predefined objective function nor prior information about the communities. By analyzing the problem of real-time community detection in large networks and having by baseline the algorithm proposed by \citep{Raghavan2007} with linear time-$O(m)$ on a network with $m$ edges-label propagation, or ``epidemic'' community detection, \citep{Leung2009} proposed a method with near linear time community detection in graphs. They identified the characteristics and drawbacks of the base \citep{Raghavan2007} algorithm and extended it by incorporating different heuristics to facilitate reliable and multifunctional real-time community detection. \citep{DBLP:journals/corr/YunLP14} proposed two efficient streaming memory-limited clustering algorithms for community detection based on spectral methods. \citep{DBLP:journals/corr/YunP14} proposed community detection via random and adaptive sampling. \citep{Sariyuce2016} proposed SONIC, a find-and-merge type of overlapping community detection algorithm that can efficiently handle streaming updates. Recently, \citep{DBLP:journals/corr/HollocouMBL17} proposed SCoDA, a linear streaming algorithm for community detection in very large networks.

\subsubsection{Density Optimization}

Modularity-based algorithms used for community detection have been increasing in recent years. Modularity and its application have been generating controversy since some authors argue it is not a metric without disadvantages. It has been shown that algorithms that use modularity to detect communities suffer a resolution limit and, therefore, it is unable to identify small communities in some situations. In this function of this package, we try to apply a density optimization of communities found by the available algorithms \citep{densopt}. We introduce a metric we call ADC (Average Density per Community); we use this metric to prove our optimization provides improvements to the community density obtained with benchmark algorithms. The results of the optimization algorithm proved to be interesting.

Several developments were made to test the hypothesis. An algorithm was developed, and a metric is introduced in the following sections.  

\paragraph{Average Density per Community (ADC) measure}

Average Density per Community (ADC) is the measure that is used to compare the algorithm results and is given by the following formula:

$$ADC = \frac{1}{n_C} \sum_{C_{i=1}}^{n} Density(Ci)$$

where $n_C$ is the number of communities identified in the graph, $Density(Ci)$ is the density of each community $Ci$.

\paragraph{Optimization Algorithm}

Algorithm \ref{densopt} provides the sequence of tasks we are doing to test the hypothesis. We start by using the results of a community detection algorithm. Then, we try to discover if the communities can be disbanded in smaller communities. These smaller communities are strongly connected components, i.e., groups of vertices with higher density. Then, if the average community density of the disbanded communities is higher than the original community the disbanding is indeed executed. If not, the community founded by the benchmark algorithm is not disbanded and maintains its original id.

\begin{algorithm}
  \caption{Algorithm Pseudo-Code for Optimization of Community Density}\label{densopt}
    \begin{algorithmic}[1]
    \renewcommand{\algorithmicrequire}{\textbf{Input:}}
    \renewcommand{\algorithmicensure}{\textbf{Output:}}
        \Require $Communities\_Data$ \Comment{Vertex List and their Community}
        \Ensure $Community\_Results$ \Comment{New Community Structure}
        
    \While{not at the end of $Original\_Communities$ list}
        \If{$ncomponents > 1$} \Comment{If community has more than 1 component}
            \State $SCC \leftarrow \Call{strong\_connected\_components\_of\_community}{Community_i}$
            \State $mdc \leftarrow \Call{mean\_density\_of\_components}{SCCs}$
            \If{$mdc > Community\_Density$}
            \For{$SCC_i \in Community$}
                
                    \State $Community\_Results \leftarrow \Call{ComponentVerticesFormNewCommunity}{SCC_i}$
            \EndFor
        \Else
                    \State $do\_nothing$
        \EndIf    
        \Else
            \State process $next\_community$
        \EndIf  
    \EndWhile
\end{algorithmic}
\end{algorithm}

\subsubsection{Python Algorithms}

\paragraph{RDyn \newline\newline}
\begin{figure*}[h]
\centering
\subfloat[]{\includegraphics[width=1\textwidth]{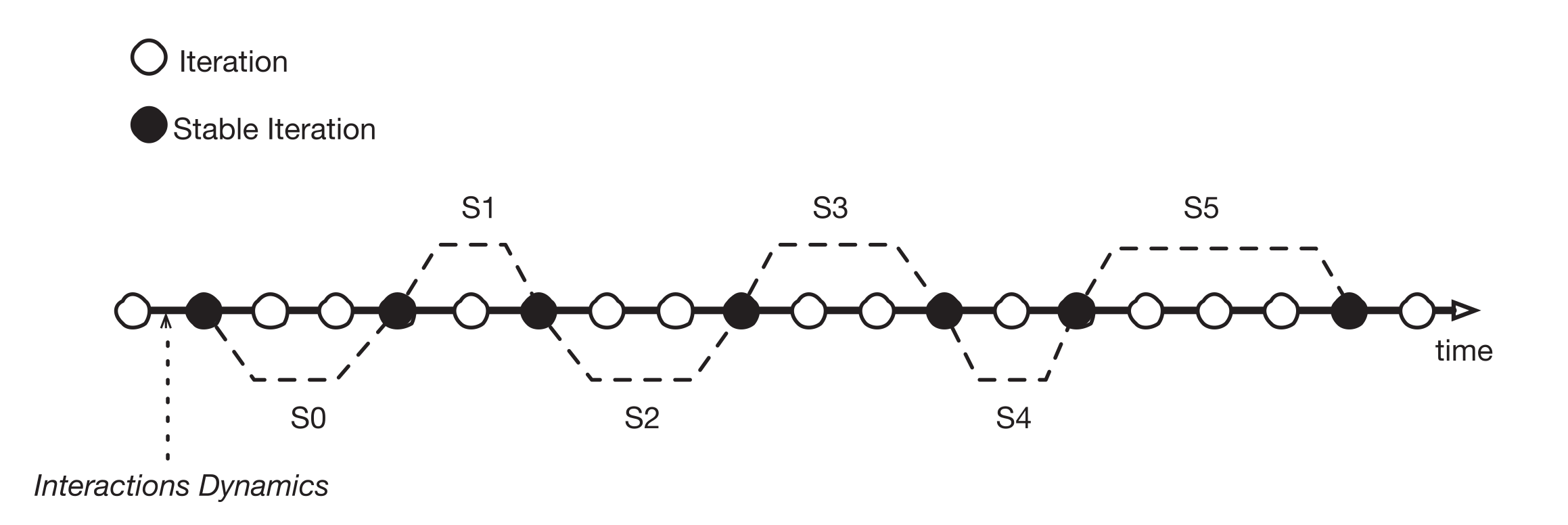}}
\caption{RDyn execution timeline: ground-truth communities are generated only during stable iterations (black circles). Interactions between two consecutive stable iterations compose a snapshot (here identified with {S0,...,S5}). Interaction dynamics (as well as community ones) happens between consecutive iteration. Due to their definitions stable iterations are not bounded to appear with fixed displacement.
}
\label{fig:RDyn}
\end{figure*}

Dynamic networks can be used to model a wide range of real-life phenomena. However, being able to access dynamic datasets having ground-truth communities is not trivial.
A classic way to address the lack of coherently annotated dataset is to employ synthetic benchmarks, such as LFR\citep{lancichinetti2008benchmark}.
For the dynamic scenario, however, only a few network generators with planted (and evolving) community structure have been proposed so far. Among them, we recognize RDyn \citep{rossetti2017graph}. 
RDyn is designed to allow its user to deeply customize the generated topology and related dynamics: it was designed to generate dynamic graphs respecting to the following characteristics: 
\begin{itemize}
    \item (i) power law vertex degree distribution; 
    \item (ii) power law community size distribution; 
    \item (iii) tunable community quality (i.e. minimum conductance, high modularity, high density...); \item (iv) edge appearance/vanishing handling; 
    \item (v) user-defined distribution for interaction decay and 
    \item (vi) communities merge/split dynamics.
\end{itemize}

As shown in Fig. \ref{fig:RDyn}, RDyn is implemented as an iterative process since the topologies it generates are the results of subsequent choices made by the vertices within it: more specifically, during every iteration the network vertices are enabled to perform a specified set of actions (i.e., create/destroy edges—all subject to specific rules). 
Moreover, once completed each iteration the status of the resulting communities is evaluated, returned if considered stable, and community dynamics are planted.

\paragraph{Tiles \newline\newline}

\begin{figure*}[h]
\centering
\subfloat[]{\includegraphics[width=.47\textwidth]{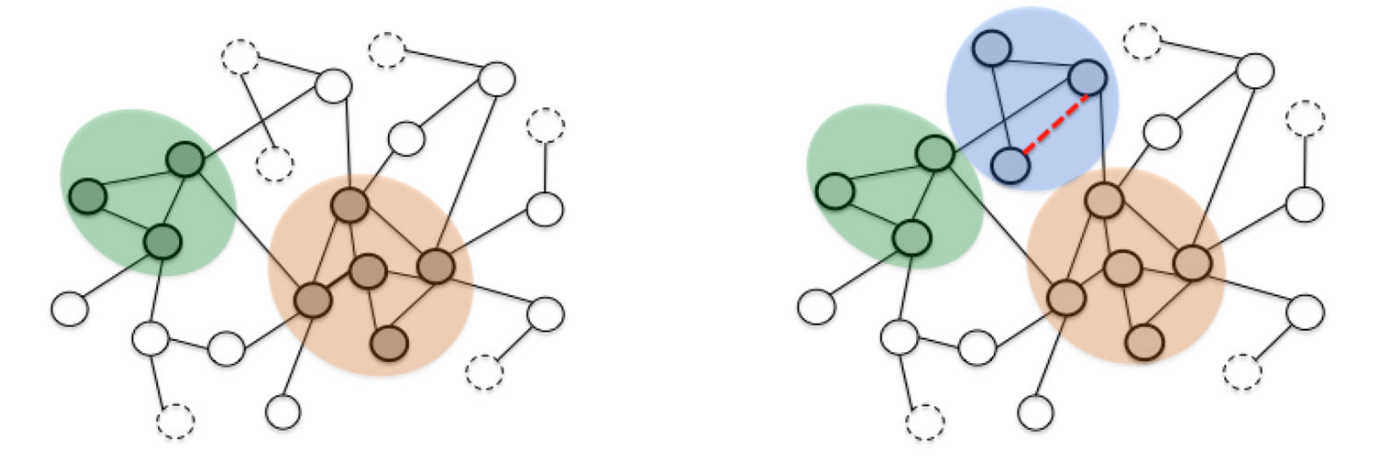}}
\subfloat[]{\includegraphics[width=.47\textwidth]{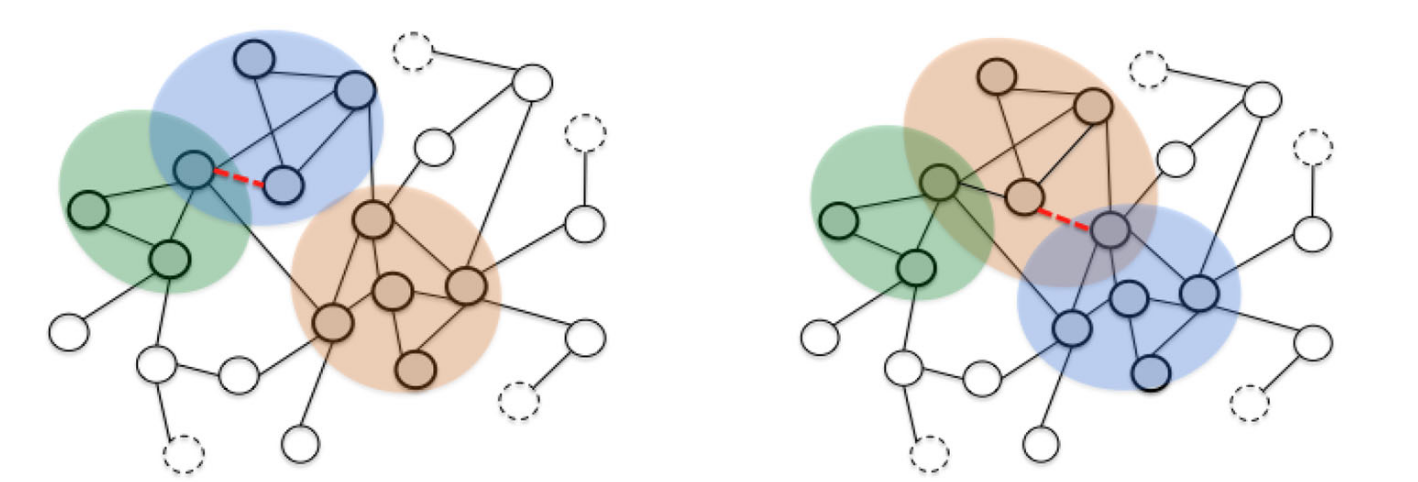}}
\caption{Example of Tiles community growth. Each new interaction is depicted with a red dashed line. Colored shapes identify core communities. Vertices with solid borders outside the colored shapes are the peripheral vertices. Vertices with dashed borders outside the colored shapes are not involved in any community.}
\label{fig:Tiles}
\end{figure*}

Social interactions determine how communities form and evolve. Indeed, the rising and vanishing of interactions can change the communities’ equilibrium.
A common approach in literature \citep{rossetti2018community} to address topology dynamics is to: 
\begin{itemize}
    \item (i) split the network into temporal snapshots; 
    \item (ii) repeat a static community detection for each snapshot and; 
    \item (iii) study the variation of the results as time goes by. 
\end{itemize}

This approach introduces an obvious issue: which temporal threshold has to be chosen to partition the network? This problem, which is context dependent, also adds another one: once the algorithm is performed on each snapshot how can we identify the same community in consecutive time slots? 
To overcome these issues, Tiles was introduced in \citep{rossetti2017tiles}: a Dynamic Community Discovery algorithm that does not impose fixed temporal thresholds for the partition of the network and the extraction of communities. 

Tiles analyzes an interaction stream and, every time a new interaction is produced by a given streaming source, it applies a label propagation procedure to diffuse the changes to the vertex surroundings
and adjust the neighbors’ community memberships (Figure \ref{fig:Tiles}). 
A vertex can belong to a community with two different levels of involvement: peripheral membership and core membership. 
Only core vertices are allowed during the label propagation phase to spread community membership to their neighbors.

\subsubsection{Cpp Algorithms}

\algorithmref{DynamicCommunityDetectionAlgorithm} presents the pseudo-code of the proposed dynamic community detection algorithm based on Louvain algorithm. The algorithm input parameters are the initial network ($G = (V,E)$) and the list of edges to be added and removed from the graph during the iterations ($A$ and $R$ respectively). For storing the community information, the algorithms use two internal community networks: the lower-level network ($\mathcal{C}_{ll}$) where the original network is maintained, and the upper-level network ($\mathcal{C}_{ul}$) where the aggregated community network is stored. The main algorithm procedure (\lineref{lst:line:Main}) handles the main flow and the several subprocedures for specific algorithm tasks. These tasks subprocedures are separated into two types. The subprocedures type that do not change the network and are used to get data from both the lower-level and upper-level  network (eg.: \texttt{CommunityChangedVertices()}, \lineref{lst:line:CommunityChangedVertices}), and subprocedures that update the networks in terms of edges or vertices and/or community assignment (eg.: \texttt{DisbandCommunities()}, \lineref{lst:line:DisbandCommunitiesAdd} or \lineref{lst:line:DisbandCommunitiesRem}). The task procedures of the algorithm are conceptual and aggregate subprocedures to complete specific tasks (i.e., Adding edge to the $\mathcal{C}_{ll}$). They are repeated until a modularity increase is possible or edges to be removed or added.

\begin{description}
    \item[Procedure P1a:] Adding edge to $\mathcal{C}_{ll}$, consists in the retrieval of a list of affected vertices and respective communities by the addition of an edge (\lineref{lst:line:AffectedByAddition}) and by the addition of the edge itself to $\mathcal{C}_{ll}$ (\lineref{lst:line:AddEdge});
    \item[Procedure P1b:] Removal of edge in the $\mathcal{C}_{ll}$. This procedure consists in the retrieval of a list of affected vertices and respective communities when the removal of an edge is performed (\lineref{lst:line:AffectedByRemoval}), and by the removal of the edge itself to $\mathcal{C}_{ll}$ (\lineref{lst:line:RemoveEdge});
    \item[Procedure P2:] Disband Affected Communities in $\mathcal{C}_{ll}$. Based on the list of affected vertices and respective communities retrieved by \texttt{AffectedByAddition()} or \texttt{AffectedByRemoval()} the affected communities will be disbanded in $\mathcal{C}_{ll}$ (\lineref{lst:line:DisbandCommunitiesAdd} or \lineref{lst:line:DisbandCommunitiesRem}); 
    \item[Procedure P3:] Update the $\mathcal{C}_{ul}$ with changes of  $\mathcal{C}_{ll}$. The list of affected vertices and respective communities retrieved by \texttt{AffectedByAddition()} or \texttt{AffectedByRemoval()} will be also used to replicate the changes in community structure to the $\mathcal{C}_{ul}$ (\lineref{lst:line:SyncCommunitiesAdd} or \lineref{lst:line:SyncCommunitiesRem}). Notice that in this procedure, the added or removed edges will also be updated in the $\mathcal{C}_{ul}$; 
    \item[Procedure P4:] $\mathcal{C}_{ul}$ will be used to perform the Louvain Algorithm Step 1 and calculate the changes in the community structure that may lead to a locally optimised modularity (\lineref{lst:line:OneLevel});     \item[Procedure P5:] Update $\mathcal{C}_{ll}$ with the communities that changed by applying the Louvain Algorithm Step 1 to $\mathcal{C}_{ul}$ (\lineref{lst:line:UpdateCommunities});   
    \item[Procedure P6:] Use the $\mathcal{C}_{ul}$ to perform the Louvain Algorithm Step 2 and aggregate communities (\lineref{lst:line:PartitionToGraph}).   
\end{description}

\section{DynComm R Package}\label{Dyncomm}

\begin{figure}[h]
    \centering
    \includegraphics[scale=0.5]{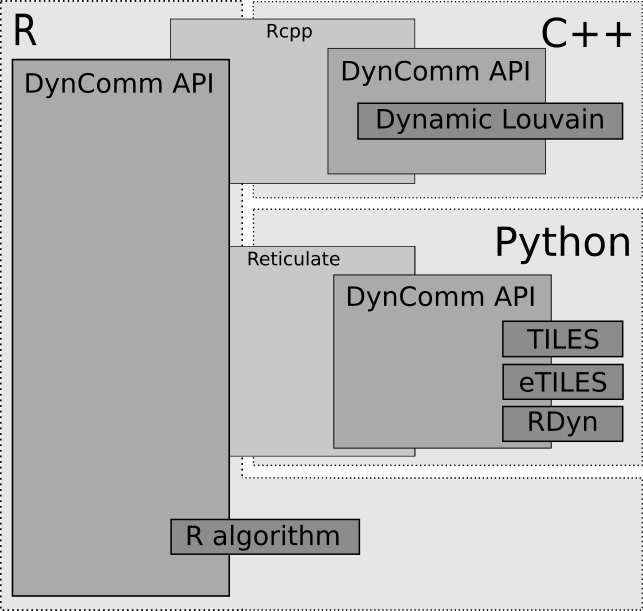}
    \caption{DynComm R package Architecture}
    \label{fig:Arch}
\end{figure}

DynComm package, although an R package, has interface with other languages. ``Rcpp” package was used to interface C++ source code with R \citep{x,x1,x2}. To make the interface with the Python language, we used the ``reticulate" package \citep{ret}. To take measurements of processing and memory use we used package "microbenchmark" \citep{z}. Figure \ref{fig:Arch} presents a block diagram of the internal setup of DynComm.

\subsection{R}

R is the main language of the package, used to develop the interface, and the bridges to other languages. Since this pacakge is related to graph analysis, some use of particular and specific packages is expected. To compute graph-based operations in R, for example, we used the ``igraph” package \citep{ww}.
In the following subsections, we will deal with the R user interface and an algorithm, that although not a community detection algorithm, might be used independently after the use of any community detection algorithm available, after detection of communities.

\subsubsection{User Interface}\label{INT}

This package tries to supply a unified, minimalist interface that is flexible enough to allow to pass any type of information to the algorithms and retrieve the results. This holds for both end users and developers of algorithms.
Since the main target of this package are community detection algorithms capable of working in a stream mode, the interface for the end user is:

\begin{itemize}
    \item a constructor that instantiates a DynComm object to hold the choice of algorithm, any required parameters, and data
    \item an addRemoveEdges function to interactively add or remove edges
    \item a function to get the resulting vertex to community mapping
    \item a function to get the quality of the current mapping
    \item and a function to get the total incremental time taken in processing
\end{itemize}

Actually, since the graph can be very big and could be cumbersome to display the entire result at a time, there are a number of auxiliary functions that allow to get the result in smaller pieces of information, like getting the communities and then getting the vertices for each individual community, one community at a time.

\subsection{Other Languages}

DynComm R package has some examples of other languages use. Included in the package, at the present version of release, we have three Python algorithms, and one C++ algorithm. Please check the following subsections for more information.

\subsubsection{Programming API}\label{API}

For algorithm developers, their implemented code must only mimic the end user interface. It must:

\begin{itemize}
    \item add their algorithm to the list of algorithms
    \item add the required parameters to the list of parameters with default values
    \item instantiate an appropriate object in the constructor of the DynComm object, where it will receive a reference to the graph along with any parameters required that were passed by the user, or default values if they were not.
    \item implement a function, or functions, to receive edges to add/remove and insert those functions inside the addRemoveEdges function
    \item optionally, implement functions to get the results, if they require extra processing before displaying. This is stated as optional because, usually, the results are taken directly from the graph and it is stored externally to the algorithm. Any additional processing must not change the result.
\end{itemize}

To help developers of algorithms implemented in languages other than R, both C++ and Python have an interface implemented, in the respective languages, that mimics the interface in R. This facilitates implementation because the developer needs to fiddle with neither Rcpp nor Python, which respectively perform the translation of C++ and Python to R.

\section{Conclusions and Future Work}\label{conc}

This document is a publication concerning the launch of the DynComm R package. It is a package developed for everyone interested in using R to do community detection in dynamic networks. The package is prepared to be improved in the future, with new algorithms added to the already interesting menu of available algorithms.

Soon, we expect to add at least a Java language algorithm. All this package code will be available in GITHUB repository \footnote{Available Code at \url{https://github.com/softskillsgroup/DynComm-R-package}.}, in development and public mode to all developers. We are expecting contributions from anyone interested in promoting their algorithm(s).

\section*{Acknowledgments}
This work was fully financed by the Faculty of Engineering of the Porto University. Rui Portocarrero Sarmento also gratefully acknowledges funding from FCT (Portuguese Foundation for Science and Technology) through a Ph.D. grant (SFRH/BD/119108/2016).

\bibliographystyle{apalike}
\bibliography{Dyncomm}

\end{document}

%% file: plots/louvain/01.tex
\resizebox{0.23\linewidth}{!}{
    \begin{tikzpicture}

        \tikzstyle{surround} = [fill=black!1,draw=black,rounded corners=3mm]
        \tikzstyle{node} = [fill=black!5]
    
        \vertex[node](l1) at (0,0) {1};
        \vertex[node](l2) at ([shift={(l1)}] -120:2.33) {2};
        \vertex[node](l3) at ([shift={(l2)}] 0:2.33) {3};
        
        \vertex[node](l5) at ([shift={(l3)}] 0:2) {5};
        \vertex[node](l4) at ([shift={(l5)}] 90:2) {4};
        \vertex[node](l6) at ([shift={(l5)}] 0:2) {6};
        \vertex[node](l7) at ([shift={(l6)}] 90:2) {7};
        
        \tikzset{EdgeStyle/.style={-, line width=0.3}}
        \Edge[label=$1$](l1)(l2)
        \Edge[label=$1$](l1)(l3)
        \Edge[label=$1$](l2)(l3)
        \Edge[label=$1$](l3)(l5)
        \Edge[label=$1$](l5)(l4)
        \Edge[label=$1$](l5)(l6)
        \Edge[label=$1$](l6)(l7)
        \Edge[label=$1$](l4)(l7);

    \end{tikzpicture}
}

%% file: plots/louvain/02.tex
\resizebox{0.23\linewidth}{!}{
    \begin{tikzpicture}

        \tikzstyle{surround} = [fill=black!1,draw=black,rounded corners=3mm]
        \tikzstyle{node} = [fill=black!25]
    
        \vertex[node](l1) at (0,0) {1};
        \vertex[node](l2) at ([shift={(l1)}] -120:2.33) {2};
        \vertex[node](l3) at ([shift={(l2)}] 0:2.33) {3};
        
        \vertex[node](l5) at ([shift={(l3)}] 0:2) {5};
        \vertex[node](l4) at ([shift={(l5)}] 90:2) {4};
        \vertex[node](l6) at ([shift={(l5)}] 0:2) {6};
        \vertex[node](l7) at ([shift={(l6)}] 90:2) {7};
        
        \tikzset{EdgeStyle/.style={-, line width=0.3}}
        \Edge[label=$1$](l1)(l2)
        \Edge[label=$1$](l1)(l3)
        \Edge[label=$1$](l2)(l3)
        \Edge[label=$1$](l3)(l5)
        \Edge[label=$1$](l5)(l4)
        \Edge[label=$1$](l5)(l6)
        \Edge[label=$1$](l6)(l7)
        \Edge[label=$1$](l4)(l7);

        %
    % Background Box
    \begin{pgfonlayer}{background}
    
    \node[surround] (background) [fit = (l1)] {};
    \node[surround] (background) [fit = (l2)] {};
    \node[surround] (background) [fit = (l3)] {};
    
    \node[surround] (background) [fit = (l4)] {};
    \node[surround] (background) [fit = (l5)] {};
    \node[surround] (background) [fit = (l6)] {};
    \node[surround] (background) [fit = (l7)] {};
   
    \end{pgfonlayer}

    \end{tikzpicture}
}

%% file: plots/louvain/03.tex
\resizebox{0.23\linewidth}{!}{
    \begin{tikzpicture}

        \tikzstyle{surround} = [fill=black!1,draw=black,rounded corners=3mm]
        \tikzstyle{node} = [fill=black!5]
    
        \vertex[node](l1) at (0,0) {1};
        \vertex[node](l2) at ([shift={(l1)}] -120:2.33) {2};
        \vertex[fill=black!25](l3) at ([shift={(l2)}] 0:2.33) {3};
        
        \vertex[fill=black!25](l5) at ([shift={(l3)}] 0:2) {5};
        \vertex[node](l4) at ([shift={(l5)}] 90:2) {4};
        \vertex[node](l6) at ([shift={(l5)}] 0:2) {6};
        \vertex[fill=black!25](l7) at ([shift={(l6)}] 90:2) {7};
        
        \tikzset{EdgeStyle/.style={-, line width=0.3}}
        \Edge[label=$1$](l1)(l2)
        \Edge[label=$1$](l1)(l3)
        \Edge[label=$1$](l2)(l3)
        \Edge[label=$1$](l3)(l5)
        \Edge[label=$1$](l5)(l4)
        \Edge[label=$1$](l5)(l6)
        \Edge[label=$1$](l6)(l7)
        \Edge[label=$1$](l4)(l7);

        %
    % Background Box
    \begin{pgfonlayer}{background}
    
    \draw [surround] ([shift={(l1)}]  60:0.75) --
                     ([shift={(l1)}]  120:0.75) --
                     ([shift={(l2)}]  180:0.75) --
                     ([shift={(l2)}]  -120:0.75) -- 
                     ([shift={(l3)}]  -60:0.75) -- 
                     ([shift={(l3)}]  0:0.75) -- cycle;
                     
    \node[surround] (background) [fit = (l4)(l5)] {};
    
    \node[surround] (background) [fit = (l7)(l6)] {};
   
    \end{pgfonlayer}

    \end{tikzpicture}
}

%% file: plots/louvain/04.tex
\resizebox{0.23\linewidth}{!}{
    \begin{tikzpicture}

        \tikzstyle{surround} = [fill=black!1,draw=black,rounded corners=3mm]
        \tikzstyle{node} = [fill=black!5]
    
        \vertex[fill=black!25](l3) at (0,0) {3};
        \vertex[fill=black!25](l5) at ([shift={(l3)}] 0:2) {5};
        \vertex[fill=black!25](l7) at ([shift={(l5)}] 0:2) {7};
        
        \tikzset{EdgeStyle/.style={-, line width=0.3}}
        \Edge[label=$1$](l3)(l5)
        \Edge[label=$2$](l5)(l7)
        \Loop[label=$6$, labelstyle=above,style={-, line width=0.3},dist=1.2cm,dir=NO](l3)
        \Loop[label=$2$, labelstyle=above,style={-, line width=0.3},dist=1.2cm,dir=NO](l5)
        \Loop[label=$2$, labelstyle=above,style={-, line width=0.3},dist=1.2cm,dir=NO](l7);

    % Background Box
    \begin{pgfonlayer}{background}
    
    \node[surround] (background) [fit = (l3)] {};
    \node[surround] (background) [fit = (l5)] {};
    \node[surround] (background) [fit = (l7)] {};
   
    \end{pgfonlayer}

    \end{tikzpicture}
}

%% file: plots/louvain/05.tex
\resizebox{0.23\linewidth}{!}{
    \begin{tikzpicture}

        \tikzstyle{surround} = [fill=black!1,draw=black,rounded corners=3mm]
        \tikzstyle{node} = [fill=black!5]
    
        \vertex[fill=black!25](l3) at (0,0) {3};
        \vertex[node](l5) at ([shift={(l3)}] 0:2) {5};
        \vertex[fill=black!25](l7) at ([shift={(l5)}] 0:2) {7};
        
        \tikzset{EdgeStyle/.style={-, line width=0.3}}
        \Edge[label=$1$](l3)(l5)
        \Edge[label=$2$](l5)(l7)
        \Loop[label=$6$, labelstyle=above,style={-, line width=0.3},dist=1.2cm,dir=NO](l3)
        \Loop[label=$2$, labelstyle=above,style={-, line width=0.3},dist=1.2cm,dir=NO](l5)
        \Loop[label=$2$, labelstyle=above,style={-, line width=0.3},dist=1.2cm,dir=NO](l7);

    % Background Box
    \begin{pgfonlayer}{background}
    
    \node[surround] (background) [fit = (l3)] {};
    \node[surround] (background) [fit = (l5)(l7)] {};
   
    \end{pgfonlayer}

    \end{tikzpicture}
}

%% file: plots/louvain/06.tex
\resizebox{0.23\linewidth}{!}{
    \begin{tikzpicture}

        \tikzstyle{surround} = [fill=black!1,draw=black,rounded corners=3mm]
        \tikzstyle{node} = [fill=black!5]
    
        \vertex[fill=black!25](l3) at (0,0) {3};
        \vertex[fill=black!25](l7) at ([shift={(l3)}] 0:2) {7};
        
        \tikzset{EdgeStyle/.style={-, line width=0.3}}
        \Edge[label=$1$](l3)(l7)
        \Loop[label=$6$, labelstyle=above,style={-, line width=0.3},dist=1.2cm,dir=NO](l3)
        \Loop[label=$8$, labelstyle=above,style={-, line width=0.3},dist=1.2cm,dir=NO](l7);

    % Background Box
    \begin{pgfonlayer}{background}
    
    \node[surround] (background) [fit = (l3)] {};
    \node[surround] (background) [fit = (l7)] {};
   
    \end{pgfonlayer}

    \end{tikzpicture}
}

%% file: plots/dlouvain/toynet05.tex
\resizebox{0.22\linewidth}{!}{
    \begin{tikzpicture}

        \tikzstyle{surround} = [fill=black!1,draw=black,rounded corners=3mm]
        \tikzstyle{node} = [fill=black!5]
    
        \vertex[node](l1) at (0,0) {1};
        \vertex[node](l2) at ([shift={(l1)}] -120:2.33) {2};
        \vertex[fill=black!25](l3) at ([shift={(l2)}] 0:2.33) {3};
        
        \vertex[fill=black!25](l5) at ([shift={(l3)}] 0:2) {5};
        \vertex[node](l4) at ([shift={(l5)}] 90:2) {4};
        \vertex[node](l6) at ([shift={(l5)}] 0:2) {6};
        \vertex[fill=black!25](l7) at ([shift={(l6)}] 90:2) {7};
        
        \tikzset{EdgeStyle/.style={-, line width=0.3}}
        \Edge[label=$1$](l1)(l2)
        \Edge[label=$1$](l1)(l3)
        \Edge[label=$1$](l2)(l3)
        \Edge[label=$1$](l3)(l5)
        \Edge[label=$1$](l5)(l4)
        \Edge[label=$1$](l5)(l6)
        \Edge[label=$1$](l6)(l7) 
        \Edge[label=$1$](l4)(l7);
        
        \tikzset{EdgeStyle/.style={dashed, -, line width=0.3}}
        \Edge(l1)(l4);
        
        %
    % Background Box
    \begin{pgfonlayer}{background}
    
    \draw [surround] ([shift={(l1)}]  60:0.75) --
                     ([shift={(l1)}]  120:0.75) --
                     ([shift={(l2)}]  180:0.75) --
                     ([shift={(l2)}]  -120:0.75) -- 
                     ([shift={(l3)}]  -60:0.75) -- 
                     ([shift={(l3)}]  0:0.75) -- cycle;
                     
    \node[surround] (background) [fit = (l4)(l5)] {};
    
    \node[surround] (background) [fit = (l7)(l6)] {};
   
    \end{pgfonlayer}

    \end{tikzpicture}
}

%% file: plots/dlouvain/toynet06.tex
\resizebox{0.22\linewidth}{!}{
    \begin{tikzpicture}

        \tikzstyle{surround} = [fill=black!1,draw=black,rounded corners=3mm]
        \tikzstyle{node} = [fill=black!5]
    
        \vertex[fill=black!25](l1) at (0,0) {1};
        \vertex[fill=black!25](l2) at ([shift={(l1)}] -120:2.33) {2};
        \vertex[fill=black!25](l3) at ([shift={(l2)}] 0:2.33) {3};
        
        \vertex[fill=black!25](l5) at ([shift={(l3)}] 0:2) {5};
        \vertex[fill=black!25](l4) at ([shift={(l5)}] 90:2) {4};
        \vertex[node](l6) at ([shift={(l5)}] 0:2) {6};
        \vertex[fill=black!25](l7) at ([shift={(l6)}] 90:2) {7};
        
        \tikzset{EdgeStyle/.style={-, line width=0.3}}
        \Edge[label=$1$](l1)(l2)
        \Edge[label=$1$](l1)(l3)
        \Edge[label=$1$](l2)(l3)
        \Edge[label=$1$](l3)(l5)
        \Edge[label=$1$](l5)(l4)
        \Edge[label=$1$](l5)(l6)
        \Edge[label=$1$](l6)(l7) 
        \Edge[label=$1$](l4)(l7)
        \Edge[label=$1$](l1)(l4);
        
        %
    % Background Box
    \begin{pgfonlayer}{background}
    
    \node[surround] (background) [fit = (l1)] {};
    \node[surround] (background) [fit = (l2)] {};
    \node[surround] (background) [fit = (l3)] {};
    
    \node[surround] (background) [fit = (l4)] {};
    \node[surround] (background) [fit = (l5)] {};
    
    \node[surround] (background) [fit = (l7)(l6)] {};
   
    \end{pgfonlayer}

    \end{tikzpicture}
}

%% file: plots/dlouvain/toynet09.tex
\resizebox{0.22\linewidth}{!}{
    \begin{tikzpicture}

        \tikzstyle{surround} = [fill=black!1,draw=black,rounded corners=3mm]
        \tikzstyle{node} = [fill=black!5]
    
        \vertex[node](l1) at (0,0) {1};
        \vertex[node](l2) at ([shift={(l1)}] -120:2.33) {2};
        \vertex[fill=black!25](l3) at ([shift={(l2)}] 0:2.33) {3};
        
        \vertex[node](l5) at ([shift={(l3)}] 0:2) {5};
        \vertex[node](l4) at ([shift={(l5)}] 90:2) {4};
        \vertex[node](l6) at ([shift={(l5)}] 0:2) {6};
        \vertex[fill=black!25](l7) at ([shift={(l6)}] 90:2) {7};
        
        \tikzset{EdgeStyle/.style={-, line width=0.3}}
        \Edge[label=$1$](l1)(l2)
        \Edge[label=$1$](l1)(l3)
        \Edge[label=$1$](l2)(l3)
        \Edge[label=$1$](l3)(l5)
        \Edge[label=$1$](l5)(l4)
        \Edge[label=$1$](l5)(l6)
        \Edge[label=$1$](l6)(l7) 
        \Edge[label=$1$](l4)(l7)
        \Edge[label=$1$](l1)(l4);
        
        %
    % Background Box
    \begin{pgfonlayer}{background}
    
    \draw [surround] ([shift={(l1)}]  60:0.75) --
                     ([shift={(l1)}]  120:0.75) --
                     ([shift={(l2)}]  180:0.75) --
                     ([shift={(l2)}]  -120:0.75) -- 
                     ([shift={(l3)}]  -60:0.75) -- 
                     ([shift={(l3)}]  0:0.75) -- cycle;
    
    \node[surround] (background) [fit = (l4)(l5)(l7)(l6)] {};
   
    \end{pgfonlayer}

    \end{tikzpicture}
}

%% file: plots/dlouvain/toynet05h.tex
\resizebox{0.22\linewidth}{!}{
    \begin{tikzpicture}

        \tikzstyle{surround} = [fill=black!1,draw=black,rounded corners=3mm]
        \tikzstyle{node} = [fill=black!5]
    
        \vertex[fill=black!25](l3) at (0,0) {3};
        \vertex[fill=black!25](l7) at ([shift={(l3)}] 0:2) {7};
        
        \tikzset{EdgeStyle/.style={-, line width=0.3}}
        \Edge[label=$1$](l3)(l7)
        \Loop[label=$6$, labelstyle=above,style={-, line width=0.3},dist=1.2cm,dir=NO](l3)
        \Loop[label=$8$, labelstyle=above,style={-, line width=0.3},dist=1.2cm,dir=NO](l7);

        \tikzset{EdgeStyle/.style={dashed, -, line width=0.3}}
        
        \draw[bend left, -, style=dashed]  (l3) to node [auto] {} (l7);
        
    % Background Box
    \begin{pgfonlayer}{background}
    
    \node[surround] (background) [fit = (l3)] {};
    \node[surround] (background) [fit = (l7)] {};
   
    \end{pgfonlayer}

    \end{tikzpicture}
}

%% file: plots/dlouvain/toynet07.tex
\resizebox{0.22\linewidth}{!}{
    \begin{tikzpicture}

        \tikzstyle{surround} = [fill=black!1,draw=black,rounded corners=3mm]
        \tikzstyle{node} = [fill=black!5]
    
        \vertex[fill=black!25](l1) at (0,0) {1};
        \vertex[fill=black!25](l2) at ([shift={(l1)}] -120:2.33) {2};
        \vertex[fill=black!25](l3) at ([shift={(l2)}] 0:2.33) {3};
        
        \vertex[fill=black!25](l5) at ([shift={(l3)}] 0:2) {5};
        \vertex[fill=black!25](l4) at ([shift={(l5)}] 90:2) {4};
        \vertex[fill=black!25](l7) at ([shift={(l5)}] 30:2) {7};
        
        \tikzset{EdgeStyle/.style={-, line width=0.3}}
        \Edge[label=$1$](l1)(l2)
        \Edge[label=$1$](l1)(l3)
        \Edge[label=$1$](l2)(l3)
        \Edge[label=$1$](l3)(l5)
        \Edge[label=$1$](l5)(l4)
        \Edge[label=$1$](l5)(l7)
        \Edge[label=$1$](l4)(l7)
        \Edge[label=$1$](l1)(l4);
        
        \Loop[label=$2$, labelstyle=above, style={-, line width=0.3 }, dist=1.2cm, dir=NO](l7);
        
        %
    % Background Box
    \begin{pgfonlayer}{background}
    
    \node[surround] (background) [fit = (l1)] {};
    \node[surround] (background) [fit = (l2)] {};
    \node[surround] (background) [fit = (l3)] {};
    
    \node[surround] (background) [fit = (l4)] {};
    \node[surround] (background) [fit = (l5)] {};
    
    \node[surround] (background) [fit = (l7)] {};
   
    \end{pgfonlayer}

    \end{tikzpicture}
}

%% file: plots/dlouvain/toynet08.tex
\resizebox{0.22\linewidth}{!}{
    \begin{tikzpicture}

        \tikzstyle{surround} = [fill=black!1,draw=black,rounded corners=3mm]
        \tikzstyle{node} = [fill=black!5]
    
        \vertex[node](l1) at (0,0) {1};
        \vertex[node](l2) at ([shift={(l1)}] -120:2.33) {2};
        \vertex[fill=black!25](l3) at ([shift={(l2)}] 0:2.33) {3};
        
        \vertex[node](l5) at ([shift={(l3)}] 0:2) {5};
        \vertex[node](l4) at ([shift={(l5)}] 90:2) {4};
        \vertex[fill=black!25](l7) at ([shift={(l5)}] 30:2) {7};
        
        \tikzset{EdgeStyle/.style={-, line width=0.3}}
        \Edge[label=$1$](l1)(l2)
        \Edge[label=$1$](l1)(l3)
        \Edge[label=$1$](l2)(l3)
        \Edge[label=$1$](l3)(l5)
        \Edge[label=$1$](l5)(l4)
        \Edge[label=$1$](l5)(l7)
        \Edge[label=$1$](l4)(l7)
        \Edge[label=$1$](l1)(l4);
        
        \Loop[label=$2$, labelstyle=above, style={-, line width=0.3 }, dist=1.2cm, dir=NO](l7);
        
        %
    % Background Box
    \begin{pgfonlayer}{background}
    
    \draw [surround] ([shift={(l1)}]  60:0.75) --
                     ([shift={(l1)}]  120:0.75) --
                     ([shift={(l2)}]  180:0.75) --
                     ([shift={(l2)}]  -120:0.75) -- 
                     ([shift={(l3)}]  -60:0.75) -- 
                     ([shift={(l3)}]  0:0.75) -- cycle;

    \draw [surround] ([shift={(l4)}]  90:0.75) --
                     ([shift={(l4)}]  150:0.75) --
                     ([shift={(l5)}]  -150:0.75) --
                     ([shift={(l5)}]  -90:0.75) -- 
                     ([shift={(l7)}]  -30:0.75) -- 
                     ([shift={(l7)}]  30:0.75) -- cycle;
   
    \end{pgfonlayer}

    \end{tikzpicture}
}

%% file: plots/dlouvain/toynet10.tex
\resizebox{0.22\linewidth}{!}{
    \begin{tikzpicture}

        \tikzstyle{surround} = [fill=black!1,draw=black,rounded corners=3mm]
        \tikzstyle{node} = [fill=black!5]
    
        \vertex[fill=black!25](l3) at (0,0) {3};
        \vertex[fill=black!25](l7) at ([shift={(l3)}] 0:2) {7};
        
        \tikzset{EdgeStyle/.style={-, line width=0.3}}
        \Edge[label=$2$](l3)(l7)
        \Loop[label=$6$, labelstyle=above,style={-, line width=0.3},dist=1.2cm,dir=NO](l3)
        \Loop[label=$8$, labelstyle=above,style={-, line width=0.3},dist=1.2cm,dir=NO](l7);

    % Background Box
    \begin{pgfonlayer}{background}
    
    \node[surround] (background) [fit = (l3)] {};
    \node[surround] (background) [fit = (l7)] {};
   
    \end{pgfonlayer}

    \end{tikzpicture}
}